\documentclass[useAMS,usenatbib,usegraphicx,footnotetext]{mn2e}

\def\k{km s$^{-1}$}
\def\ks{km s$^{-1}$~}

\def\m{$^\prime$}
\def\s{$^{\prime\prime}$}

\def\cm3{cm$^{-3}$}

\def\2{$^{12}$CO}
\def\3{$^{13}$CO}
\def\msol{M$_\odot$}

\newcommand*\samethanks[1][\value{footnote}]{\footnotemark[#1]}

\title[The SNR G18.1-0.1 and neighboring HII regions]
  {The interstellar medium and the massive stellar content toward the SNR G18.1-0.1 and neighboring HII regions} 

\author[S. Paron, W. Weidmann, M. E. Ortega, J. F. Albacete Colombo, and A. Pichel]{S. Paron$^{1,2,3}$\thanks{E-mail:
sparon@iafe.uba.ar}\thanks{Visiting Astronomer, Complejo Astron\'omico El Leoncito operated
under agreement between the Consejo Nacional de Investigaciones Cient\'\i ficas y
T\'ecnicas de la Rep\'ublica Argentina and the National Universities of La Plata, C\'ordoba and San Juan.}, W. Weidmann$^{4}$\samethanks, 
M. E. Ortega$^{1}$, J. F. Albacete Colombo$^{5}$ and A. Pichel$^{1}$\\
$^{1}$ Instituto de Astronom\'\i a y  F\'\i sica del Espacio
(CONICET-UBA), CC 67, Suc. 28, 1428 Buenos Aires, Argentina\\
$^{2}$ FADU, Universidad de Buenos Aires, Ciudad Universitaria, Buenos Aires, Argentina\\
$^{3}$ CBC, Universidad de Buenos Aires, Ciudad Universitaria, Buenos Aires, Argentina\\
$^{4}$ Observatorio Astron\'omico C\'ordoba, Universidad Nacional de C\'ordoba, Argentina\\ 
$^{5}$ Centro Regional Zona Atl\'antica (CURZA) - Universidad Nacional del Comahue, Viedma, Argentina\\
}  

\begin{document}

\date{Accepted XXXX. Received XXXX; in original form XXXX}

\pagerange{\pageref{firstpage}--\pageref{lastpage}} \pubyear{2012}

\maketitle

\label{firstpage}

\begin{abstract}

We perform a multiwavelength study toward the SNR G18.1-0.1 and nearby several HII regions (infrared dust bubbles N21 and N22, and the
HII regions G018.149-00.283 and G18.197-00.181). Our goal is to provide observational evidence supporting that massive stars usually
born in clusters from the same molecular cloud, which then produce, along their evolution, different neighboring objects such as HII regions, 
interstellar bubbles and supernova remnants.
We suggest that the objects analysed in this work belong to a same complex located at the distance of about 4 kpc.
Using molecular data we inspected the interstellar medium toward this complex and from optical and X-ray observations we looked for OB-type stars in the region.
Analysing public $^{13}$CO J=1--0 data we found several molecular structures very likely related to
the HII region/SNR complex. We suggest that the molecular gas is very likely being swept and shaped by the expansion of the HII regions. 
From spectroscopic optical observations obtained with the 2.15 m telescope at CASLEO, Argentina, we discovered three O-type stars very likely exciting 
the bubbles N21 and N22, and an uncatalogued HII region northward bubble N22, respectively. Also we found four B0-5 stars, one toward the bubble N22 
and the others within the HII region G18.149-0.283. By inspecting the Chandra Source Catalog we found two point X-ray sources and we suggest that one 
of them is an early O-type star. 
Finally we inspected the large scale interstellar medium around this region. We discovered a big molecular
shell of about 70 pc $\times$ 28 pc in which the analysed complex appears to be
located in its southern border.

\end{abstract}


\section{Introduction}

\begin{figure}
\centering
\includegraphics[width=9.6cm]{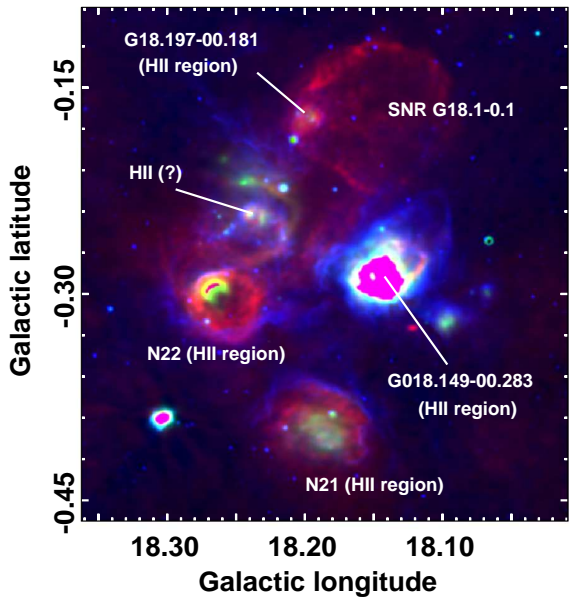}
\caption{The studied region presented in a three-colour image. The radio continuum emission at 20 cm
is displayed in red, the IRAC-{\it Spitzer} 8 $\mu$m emission in blue and the MIPS-{\it Spitzer} 24 $\mu$m emission is shown in green. 
The objects analysed in this work are indicated. The arc feature in N22 and
almost the whole interior of G18.149-0.283 are saturated in 24 $\mu$m.}
\label{presentfig}
\end{figure}

Nowadays it is well known that massive stars in our Galaxy are born predominantly within the dense cores of giant 
molecular clouds (for a thorough discussion of the star-formation theories see \citealt{mckee07}). They usually 
form and evolve in clusters, hence it is expectable to observe several HII regions in different evolutionary stages and probably 
also supernova remnants (SNRs) in a same Galactic neighborhood. Some examples of this spatial coincidence
are the field toward W28, with several SNRs and HII regions (e.g. \citealt{nich11}), the SNR W44 and 
the HII region G034.8-0.7 \citep{paron09,ortega10}, the luminous blue variable (LBV) star candidate G24.73+0.69, 
some HII regions and the SNR G24.7+0.6 \citep{alber11,alber12}.
Moreover, it is usually observed large amounts of molecular gas in the surroundings of HII regions and SNRs. 
The shock and ionization fronts from these objects, which compress and sweep up the molecular gas, can 
trigger the formation of a new generation of massive stars. This has been extensively probed towards the borders of 
HII regions (e.g. \citealt{deha05,poma09,brand11}) where processes such as ``collect and collapse'' \citep{elme77} and/or 
radiative driven implosion \citep{lefloch94,lefloch95} can be taking place.

The goal of this work is to provide observational evidence supporting that massive stars usually
born in clusters from material of the same molecular cloud, which then produce, along their evolution, neighboring HII regions,
interstellar bubbles and SNRs that can interact with the parental cloud. 
To perform this we select a very rich region in which lie the SNR G18.1-0.1 and several HII regions, two of them catalogued as 
infrared dust bubbles by \citet{church06}. In Section \ref{present}, we describe the studied HII region/SNR complex, 
Section \ref{ism} presents an analysis of the interstellar medium (ISM) around these sources, in Section \ref{obstars} 
we present an optical spectroscopic search for OB-type stars (the HII regions 
exciting star candidates) and an analysis of two X-ray point sources, Section \ref{larger} presents a larger scale ISM study around the complex, 
and finally Section \ref{summ} summarizes the obtained results.

\section{The studied HII region/SNR complex}
\label{present}

In Figure \ref{presentfig} we present the 22\m~$\times$ 22\m~region where the HII region/SNR complex lies in a colour composite image as seen in 
the {\it Spitzer}-IRAC 8 $\mu$m emission (blue), the radio continuum emission at 20 cm (red), and the the {\it Spitzer}-MIPSGAL emission at 24 $\mu$m (green).
We used the mosaicked image from the  Galactic Legacy Infrared Mid-Plane Survey Extraordinaire (GLIMPSE) in the {\it Spitzer}-IRAC 
band at 8 $\mu$m which has an angular resolution of $\sim$1\farcs9 (see \citealt{benjamin03}).
MIPSGAL is a survey of the same region as GLIMPSE, using MIPS instrument (24 and 70 $\mu$m) on {\it Spitzer} \citep{carey09}.
The MIPSGAL resolution at 24 $\mu$m is 6\s. The radio continuum data at 20 cm, with a FWHM synthesized beam of about 5\s, was extracted from the
New GPS of the Multi-Array Galactic Plane Imaging Survey \citep{helfand06} which was conducted using the Very Large Array (VLA).
In what follows we describe each object based on the information that appears in the literature.

\subsection{SNR G18.1-0.1}

SNR G18.1-0.1 or G18.16-0.16 \citep{green09} is a poor studied supernova remnant that was firstly observed at 57.5 MHz by \citet{ode86}, who
identified it as SNR G18.1-0.2. The author 
also mentions the presence of the source Sharpless 2-53 \citep{sharp59}, which is related to the HII regions complex composed by
N21, N22 and G018.149-00.283 as seen in Figure \ref{presentfig}. Based on 
the distance estimated for the HII regions complex of 4.5 kpc \citep{downes80}, \citet{ode86} suggests this value as the
minimum kinematic distance for the SNR. Later, the SNR G18.1-0.1 was associated with the X-ray source AXJ182435-1311,
which was observed with ASCA \citep{sugizaki}. More recently, this shell SNR was identified in the Very Large Array survey of the Galactic plane 
at 90 cm by \citet{brogan06}, reporting fluxes of 7.6, 3.9, and 3.0 Jy at 90, 20, and 11 cm, respectively. They derived 
an spectral index of $\alpha=-0.5~(S_{\nu} \propto \nu^{\alpha})$ using the emission at 90 and 20 cm, and $\alpha=-0.4$ using the 90 and 11 cm
data. 

\subsection{Infrared dust bubbles N21 and N22}

\citet{church06,church07} using GLIMPSE data, catalogued almost 600 infrared dust
bubbles: full or partial rings bordered by a photodissociation region (PDR), seen mainly at 8 $\mu$m,
which usually encloses ionized gas and hot dust observed at 24 $\mu$m. Most of these bubbles are HII regions as is the 
case of bubbles N21 and N22  displayed in Figure \ref{presentfig}. The bubble N22, G18.259-0.307 HII region 
\citep{kolpak03}, has a recombination line velocity of v$_{\rm LSR}$ $\sim 51$ \k, 
and is located at $\sim$ 4.1 kpc. The interstellar medium around N22 was very recently investigated by \citet{ji12}, 
finding some evidence of triggered star formation.
The bubble N21 is not catalogued in \citet{kolpak03} but it appears as the HII region G018.20-00.40 in \citet{lockman89} 
with a systemic velocity of  v$_{\rm LSR}$ $\sim 43.2$ \k.
The spectral resolutions in \citet{kolpak03} and \citet{lockman89} are  2.5 \k~and between 2 and 4 \k, respectively.

\citet{watson08} identified 21 YSO candidates around N21, and based on a SED fitting to numerical hot stellar photosphere models, 
they suggested some stars to be the possible exciting sources of this HII region. 
According to \citet{anderson09}, who analysed a sample of 291 Galactic HII regions with the aim of resolving the distance
ambiguity, N21 is related to the source called C18.19-0.40 with a resolved distance d$=$3.6 kpc, and N22 to C18.26-0.30 with d$=$4.0 kpc.

\subsection{HII region G018.149-00.283}

HII region G018.149-00.283 is not catalogued as an infrared dust
bubble, and according to \citet{kolpak03}, has a recombination line velocity of  v$_{\rm LSR}$ $\sim 53.9$ \ks and is located at 
the distance of $\sim$ 4.1 kpc. This is in agreement with the distance catalogued for U18.15-0.28, the source related to 
this HII region in \citet{anderson09}

\subsection{HII region G18.197-00.181}

The source that is superimposed over the eastern border of the SNR is the HII region 
G18.197-00.181 \citep{lockman89}, which has a systemic velocity  v$_{\rm LSR}$ $\sim 46.1$ \k.
This HII region is related to C18.20-0.18 in \citet{anderson09}, whose distance (d$=$3.7 or 12.4 kpc) is not
resolved because it presents a discrepancy between the used methods to resolve the ambiguity.

\subsection{HII region (?)}

In Figure \ref{presentfig}, based on the morphology of the IR and radio continuum emission, we indicated another
possible HII region (it appears with an interrogation mark in the figure), which was not found in any catalogue.
This source has a PDR visible in the 8 $\mu$m band, which encloses ionized gas observed at 20 cm and hot dust observed at 24 $\mu$m,
typical characteristics of HII regions (see e.g. \citealt{povich07,watson08,everett10}). In this case it can 
be appreciated two peaks at 24 $\mu$m, one toward the center and the other almost in a border of this possible HII region.
According to the coordinates, this object should be HII region G18.237-0.240.

In order to summarize, Table \ref{sources} presents the sources, their Galactic coordinates, the radius in the case of the SNR, the average 
radius for the HII regions and the linear sizes. The SNR radius was extracted 
from \citet{green09},  while the average radius of bubbles N21 and N22 from \citet{church06}. In the case of G018.149-00.283, G18.197-00.181 
and the possible unknown HII region, the average radius was estimated as was done by \citet{church06} for the infrared dust bubbles. The linear
size is the spatial diameter of the sources by assuming a distance of 4 kpc.

\begin{table}
\caption{Analysed sources.}
\centering
\begin{tabular}{lcccc}
\hline
Source & $l$ & $b$ & radius & linear size (pc) \\
\hline
SNR G18.1-0.1   & 18\fdg147  & -0\fdg169 & 8\farcm0 & 9.3    \\
G18.197-00.181  & 18\fdg197  & -0\fdg181 & 0\farcm5 & 0.6   \\
G018.149-00.283   & 18\fdg149  & -0\fdg283 & 1\farcm6 & 1.8   \\
N21             & 18\fdg190  & -0\fdg396 & 2\farcm2 & 2.5  \\
N22             & 18\fdg254  & -0\fdg305 & 1\farcm7 & 1.9  \\
HII(?)          & 18\fdg237  & -0\fdg240 & 0\farcm5 & 0.6  \\
\hline
\label{sources}
\end{tabular}
\end{table}

\section{The interstellar medium}
\label{ism}

\begin{figure*}
\centering
\includegraphics[width=11.5cm]{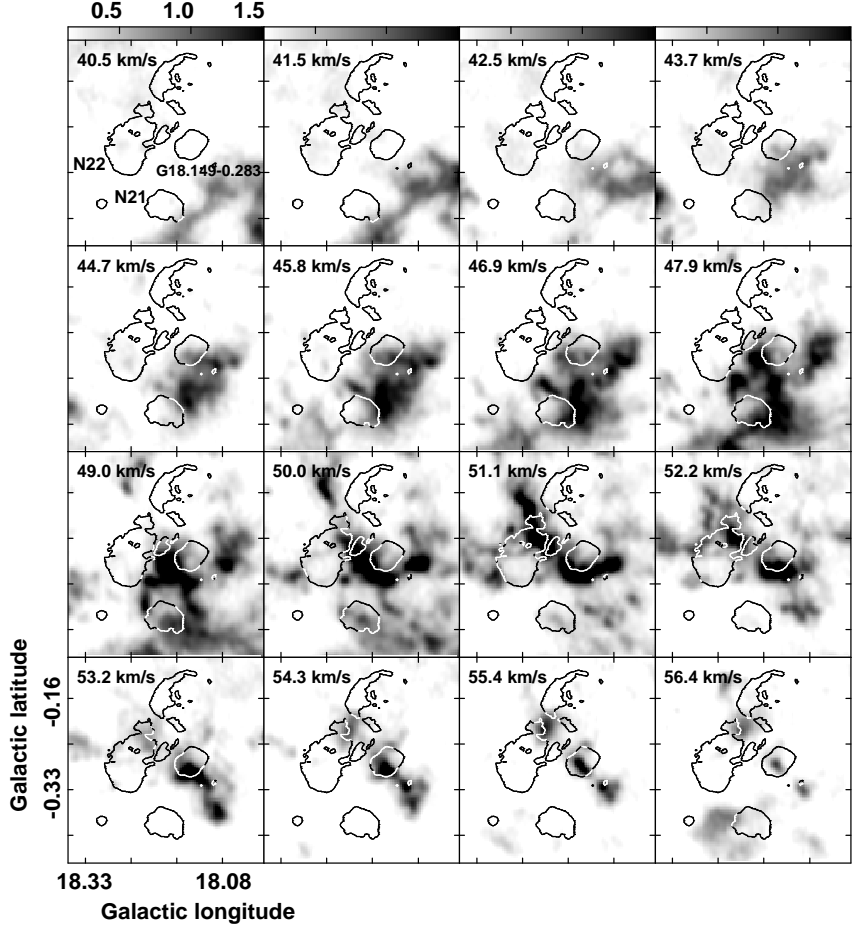}
\caption{Integrated velocity channel maps of the \3 J=1--0 emission (in grey) every $\sim 1$ \k. The greyscale is displayed
at the top of the figure and is in K \k. The contours are the smoothed radio continuum emission. The HII region G18.149-0.283 and
bubbles N21 and N22 are indicated in the first panel.  }
\label{panels}
\end{figure*}

\citet{anderson09b} analysed individually the molecular content toward a sample of 301 Galactic HII regions, which contains the HII regions
studied in this work. Using  the same molecular database we analyse the molecular material toward the whole SNR/HII region complex in order to 
have a complete picture of the relation between the molecular gas and the complex. The data were extracted from the
Galactic Ring Survey (GRS) which was performed by the Boston University and the
Five College Radio Astronomy Observatory. The survey maps the Galactic Ring in the \3 J=1--0 line
with an angular and spectral resolution of 46\s~and 0.2 \k, respectively (see \citealt{jackson06}).
The observations were performed in both position-switching and On-The-Fly mapping modes, achieving an
angular sampling of 22\s.

\begin{figure}
\centering
\includegraphics[width=9.2cm]{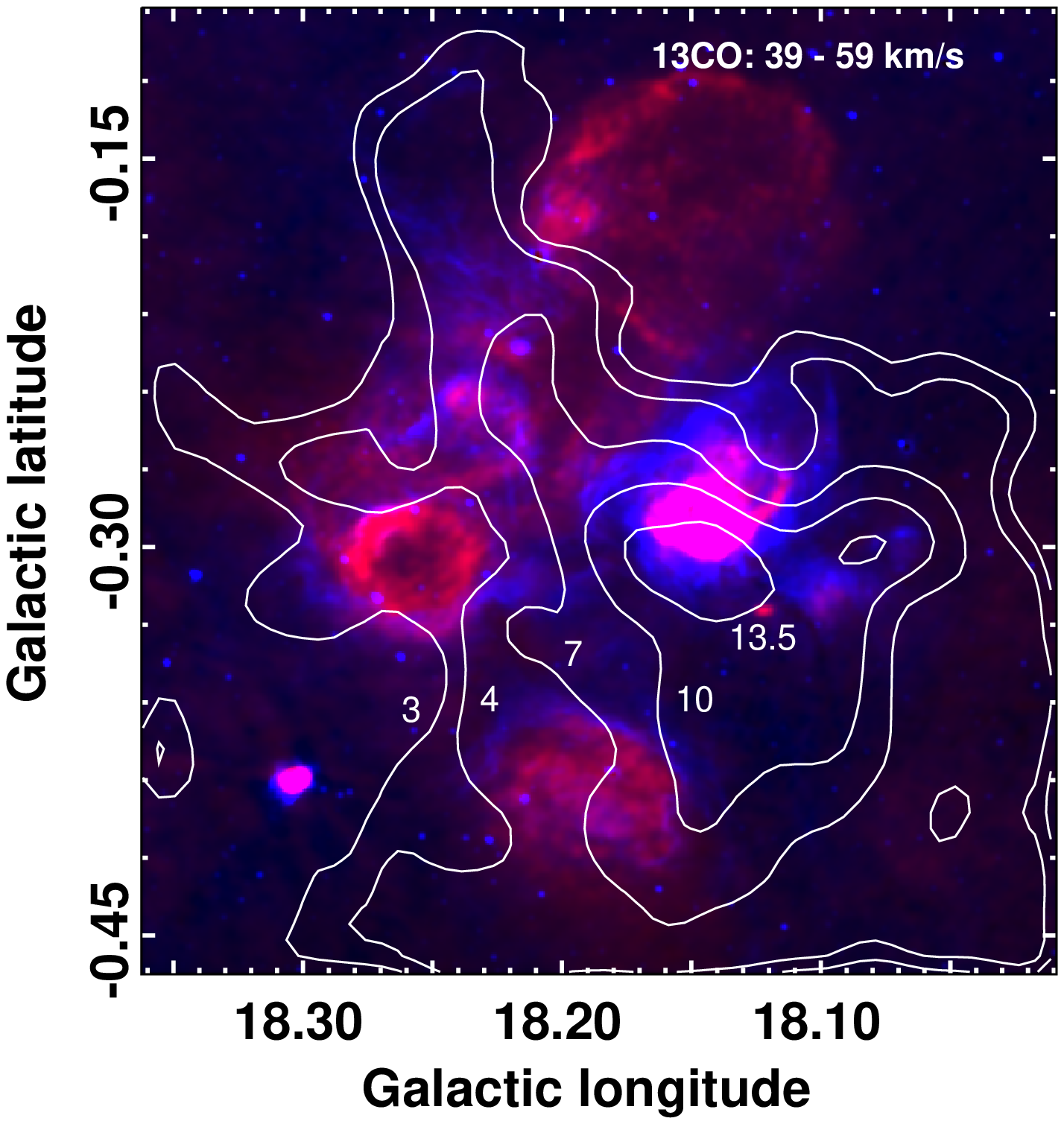}
\caption{Two-colour image with the radio continuum emission at 20 cm displayed in red and the IRAC-{\it Spitzer} 8 $\mu$m
emission in blue. The white contours are the \3 J=1--0 emission integrated from 39 to 59 \ks with levels of 3, 4, 7, 10, and 13.5 K \ks  
(also indicated in the image). The rms noise of the \3 integrated emission is about 0.5 K \k.}
\label{integ}
\end{figure}

By inspecting the whole \3 J=1--0 data cube toward the HII region/SNR complex we find several molecular structures very likely related to 
the sources within the velocity range that goes from 40 to 57 \k. Figure \ref{panels} displays the integrated velocity channel maps of 
the \3 J=1--0 emission every $\sim$ 1 \ks shown in grey. The contours are the smoothed radio
continuum emission, which are included just to mark the sources position. For a better localization of them, in the first panel, 
the HII region G18.149-0.283 and bubbles N21 and N22 are indicated. The molecular gas is mostly distributed in the surroundings of bubbles N21, N22 
and G18.149-0.283, appearing in the interstices between them, which indicates that it is very likely being swept and shaped by their expansion. 
It can be appreciated that  N21 (v$_{\rm LSR} \sim 43$ \k) is surrounded by molecular gas mainly in the velocity range  42.6 -- 52.4 \k, 
the N22 bubble (v$_{\rm LSR} \sim 51$ \k) has likely associated molecular gas between 47.9 and 54.3 \ks (for a deeper analysis of the ISM 
around this bubble see \citealt{ji12}), and 
in the velocity range that goes from 45.8 to 56.4 \ks appears gas likely related to  G018.149-00.283 
(v$_{\rm LSR} \sim 54$ \k). Concerning to the SNR and the HII region G18.197-00.181, it can be noticed that some molecular gas 
appears toward the SNR eastern and southern borders. This can be better appreciated in Figure \ref{integ}, which shows the contours 
of the \3 J=1--0 emission integrated from 39 to 59 \ks over the radio continuum at 20 cm and the IR at 8 $\mu$m emissions. It is noticeable
how the lower \3 J=1--0 contour surrounds the southeastern borders of the SNR and HII region G18.197-00.181, suggesting a 
relation between these objects and the molecular gas. Moreover, some molecular gas appears entering into the SNR radio continuum shell, 
precisely toward the south of the HII region G18.197-00.181 where the radio continuum emission is weaker.

The velocity range in which appears the molecular gas very likely related to the complex is in agreement with the
catalogued systemic velocities of the HII regions. From the Galactic rotation model of \citet{fich89}, this 
velocity range gives the near distance interval of about 3.6 -- 4.5 kpc and the farthest distance range of about 11.8 -- 12.7 kpc. 
The near distance interval is in agreement with the distances 
catalogued in \citet{anderson09}. Moreover, taking into account that
the detection rate for IR dust bubbles in GLIMPSE peaks at the
distance of 4.2 kpc within a horizon of 8 kpc \citep{church06}, we can confirm the near distance for the bubbles N21 and N22. 
Hereafter we assume a distance of about 4 kpc for the whole complex.

In order to have a rough estimate of the mass of the molecular gas we use the
\3 J=1--0 emission and assume local thermodynamic equilibrium (LTE) to obtain the H$_{2}$ column density
toward the molecular cloud shown in Figure \ref{integ}. We use:
\begin{equation}
$${\rm N(^{13}CO)} = 2.42 \times 10^{14} \frac{T_{\rm ex} \int{\tau_{13} dv}}{1 - exp(-5.29/T_{\rm ex})} \label{eq1}
\end{equation}
to obtain the \3 column density, where $T_{\rm ex}$ is the excitation temperature and $\tau_{13}$ the 
optical depth. Assuming that the $^{13}$CO J=1--0 line is optically thin, we use the approximation
\begin{equation}
\int{\tau_{13} dv} \sim \frac{1}{J(T_{ex}) - J(T_{BG})} \int{T_{\rm B} dv} \label{eq2}
\end{equation}
where
\begin{equation}
J(T) = \frac{5.29}{e^{5.29/T} - 1}, \label{eq3}
\end{equation}
$T_{BG} = $ 2.7 K is the background temperature and $T_{\rm B}$ is the brightness temperature of the line. 
Following \citet{anderson09b} we assume that $T_{\rm ex}$ is 20 K. The integration in Equation (\ref{eq2}) was done between 39 and 59 \k.
Using the relation N(H$_{2}$)/N(\3)$ \sim 5 \times 10^5$ (e.g. \citealt{simon01}) 
we obtain an averaged column density of N(H$_{2}$) $\sim 7 \times 10^{21}$ cm$^{-2}$ for the structure delimited by the 3 K \ks contour in 
Figure \ref{integ}. 
Finally we estimate a mass of $\sim 10^{5}$ \msol~for the molecular cloud. This value
was obtained from:
\begin{equation}
{\rm M} = \mu~m_{{\rm H}} \sum_{i}{\left[ D^{2}~\Omega_{i}~{\rm N_{i}(H_{2})} \right] }, \label{eq4}
\end{equation}
where $\Omega$ is the solid angle subtended by the \3 J=1--0 beam size, $m_{\rm H}$ is the hydrogen mass,
$\mu$ is the mean molecular weight that is assumed to be 2.8 by taking into account a relative helium abundance
of 25 \%, and $D$ is the distance assumed to be 4 kpc. Summation was performed over the whole molecular structure, i.e. over all 
the observed positions within the 3 K \ks contour level (see Figure \ref{integ}).

Additionally, we estimate the molecular density of the cloud densest portion, i.e. the area delimited by the 13.5 K \ks contour in Figure \ref{integ}. By 
considering an ellipsoid of semi-axis of 2\m $\times$ 1\m~centered at $l=$18\fdg146, $b=-$0\fdg308, we obtain a mass of $5.5 \times 10^{3}$ \msol, which
gives a density of about $9 \times 10^{3}$ cm$^{-3}$. 
The estimated mass of the whole molecular cloud and the density of its densest portion show that there is enough gas to probably form another
generation of massive stars. Thus, we conclude that this region is a good scenario to test triggered star formation processes in further works.

\section{Looking for OB-type stars}
\label{obstars}

\subsection{Optical observations}

\begin{figure}
\centering
\includegraphics[width=7.5cm, angle=-90]{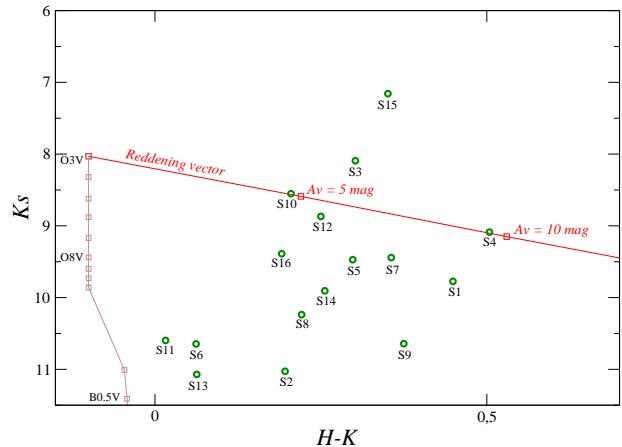}
\caption{Colour-magnitude Ks vs. (H-Ks)  diagram of the selected OB star
candidates toward the complex. The source numbers are the same as presented in Figure \ref{exciting}. The
squares along the vertical lines indicate the location of the no-reddened  main sequence stars
between O3V and B0.5V using a distance of 4 kpc. The reddening slope for
an O3V \citep{rieke85} is shown with a red line with the squares placed at
intervals corresponding to five magnitudes of visual extinction.
 }
\label{CM}
\end{figure}

\begin{figure}
\centering
\includegraphics[width=9cm]{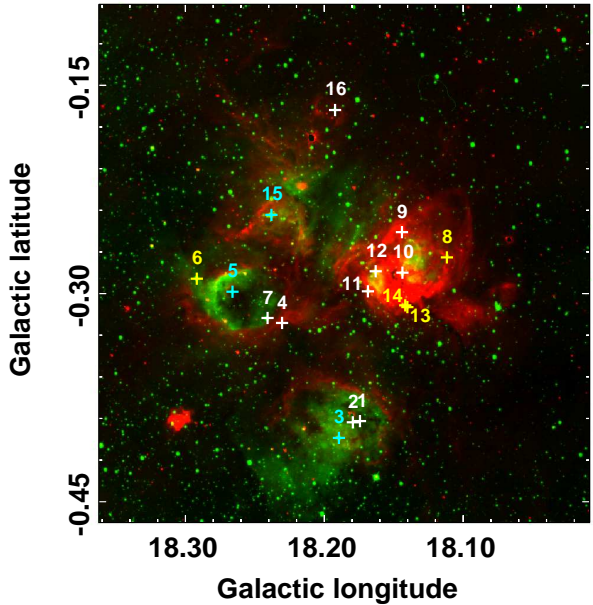}
\caption{Two-colour image with the IRAC-{\it Spitzer} 8 $\mu$m emission displayed in red, and the H$\alpha$ from the SuperCOSMOS H-alpha 
Survey \citep{halpha}, in green. The crosses show the position of the stars that were successfully observed with the 2.15\,m telescope at CASLEO, 
Argentina. They are numbered according to Table \ref{starsTable}. The discovered O an B-type stars are displayed in cyan and yellow, respectively.}
\label{exciting}
\end{figure}

In order to look for sources responsible of ionizing the gas in the HII regions, we firstly search for OB-type in catalogues such as 
\citet{reed03}. Besides the stars suggested by \citet{watson08} to ionize the N21 bubble, we did not find any OB star catalogued  
toward the whole region displayed in Figure \ref{presentfig}.
Thus, with the goal of finding the HII region exciting sources (or some of them) and possible nearby massive stars, we perform
spectroscopic optical observations toward a sample of stars. We focus our attention on those stars lying inside a region defined
by the external radius of each HII region, i.e. the external border delimited by the 8 $\mu$m emission. 
Then, we select the stars whose position in a typical near-IR colour-magnitude diagram (Ks vs. (H-Ks)) suggests that they 
are OB-type stars. This diagram was constructed by using the 
magnitudes extracted from the 2MASS All-Sky Point Source Catalog and by assuming a
distance of about 4 kpc.
Finally, we select those stars that can be observed in optical wavelengths with the 2.15\,m 
telescope at CASLEO, Argentina, i.e., the optical counterparts of the 2MASS sources with a limit of V=15.6 mag according to the 
Naval Observatory Merged Astrometric Dataset (NOMAD; \citealt{zacha04}).
Figure \ref{CM} shows the position of the selected sources in the near-IR colour-magnitude diagram, and Figure \ref{exciting} presents the physical 
position of these stars, successfully observed at CASLEO, over a two-colour image (8 $\mu$m = red, H$\alpha$ = green). As can be appreciated
in Figure \ref{CM} all sources are located in the region of OB-type stars with exception of S15. However we decided to observe it in CASLEO 
because it lies at the center of the possible unknown HII region.

\subsubsection{Observations and data reduction}

The optical observations were carried out in two runs in 2011 and 2012.
The REOSC spectrograph attached to the 2.15\,m
telescope at CASLEO was used in simple dispersion.
The spectra were taken with gratings of
300 and 600~grooves mm$^{-1}$, providing a dispersions of
3.4 and 1.6~\AA\,px$^{-1}$~with wavelength ranges of
$3500$--$7000$~\AA \ and $5220$--$6830$~\AA, respectively.
The slit was opened to 3\farcs0 and 2\farcs5
(consistent with the seeing at the site).
The exposure time was 3600 secs. in all cases.
The optical spectra were reduced and normalised using
standard techniques with
IRAF\footnote{IRAF: the Image Reduction and Analysis
Facility is distributed by the National Optical
Astronomy Observatories, which is operated by the
Association of Universities for Research in Astronomy,
Inc. (AURA) under cooperative agreement with the
National Science Foundation (NSF).}.

In general, the obtained spectra have good S/N ratios and few absorption lines were identified.
They are useful to perform a rough spectral type classification of the observed stars. At least,
O and/or B-type stars can be distinguished from other spectral types, which is important to point the presence
of massive and ionizing stars in the region.

\begin{figure}
\centering
\includegraphics[width=8.6cm]{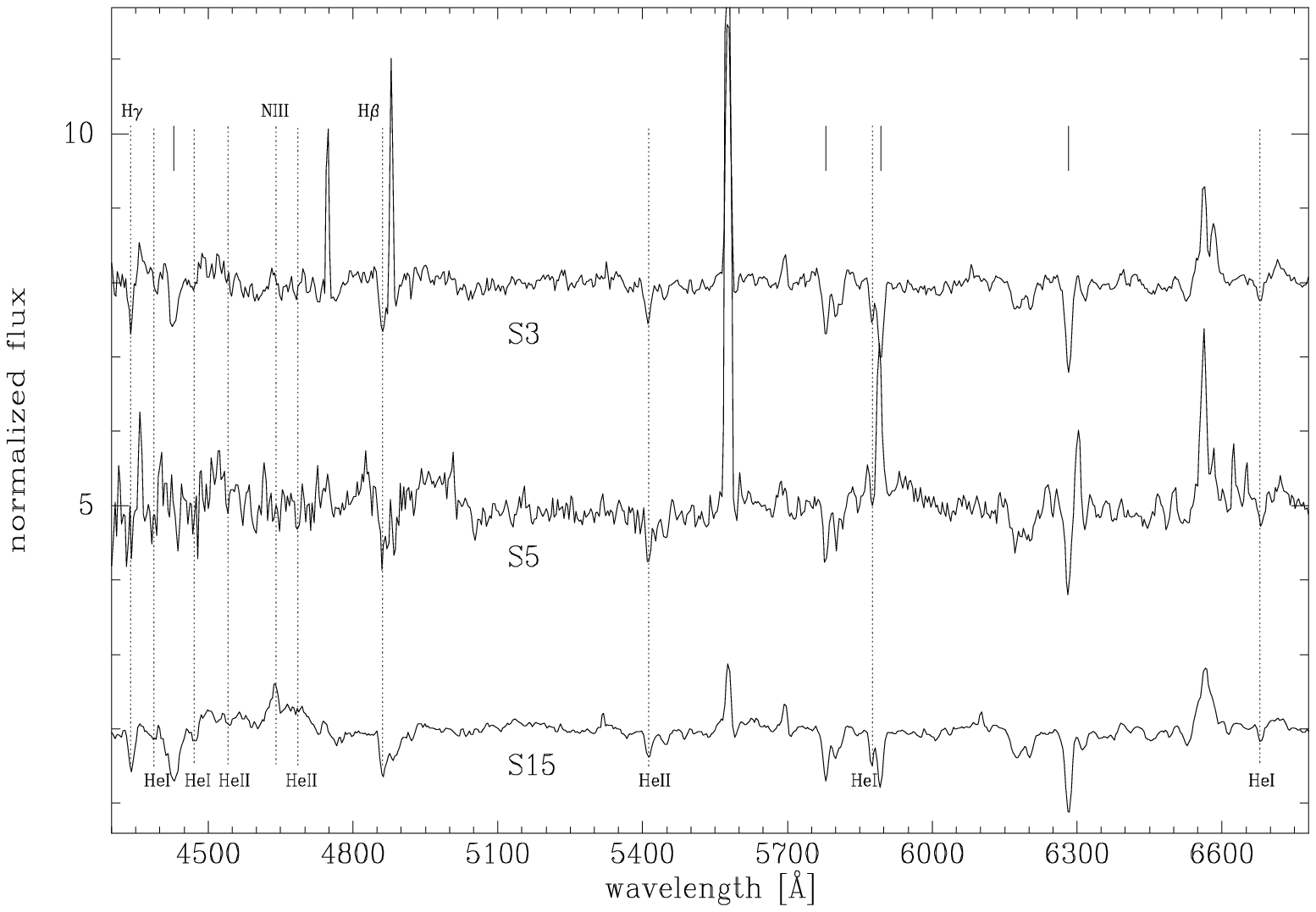}
\includegraphics[width=8.6cm]{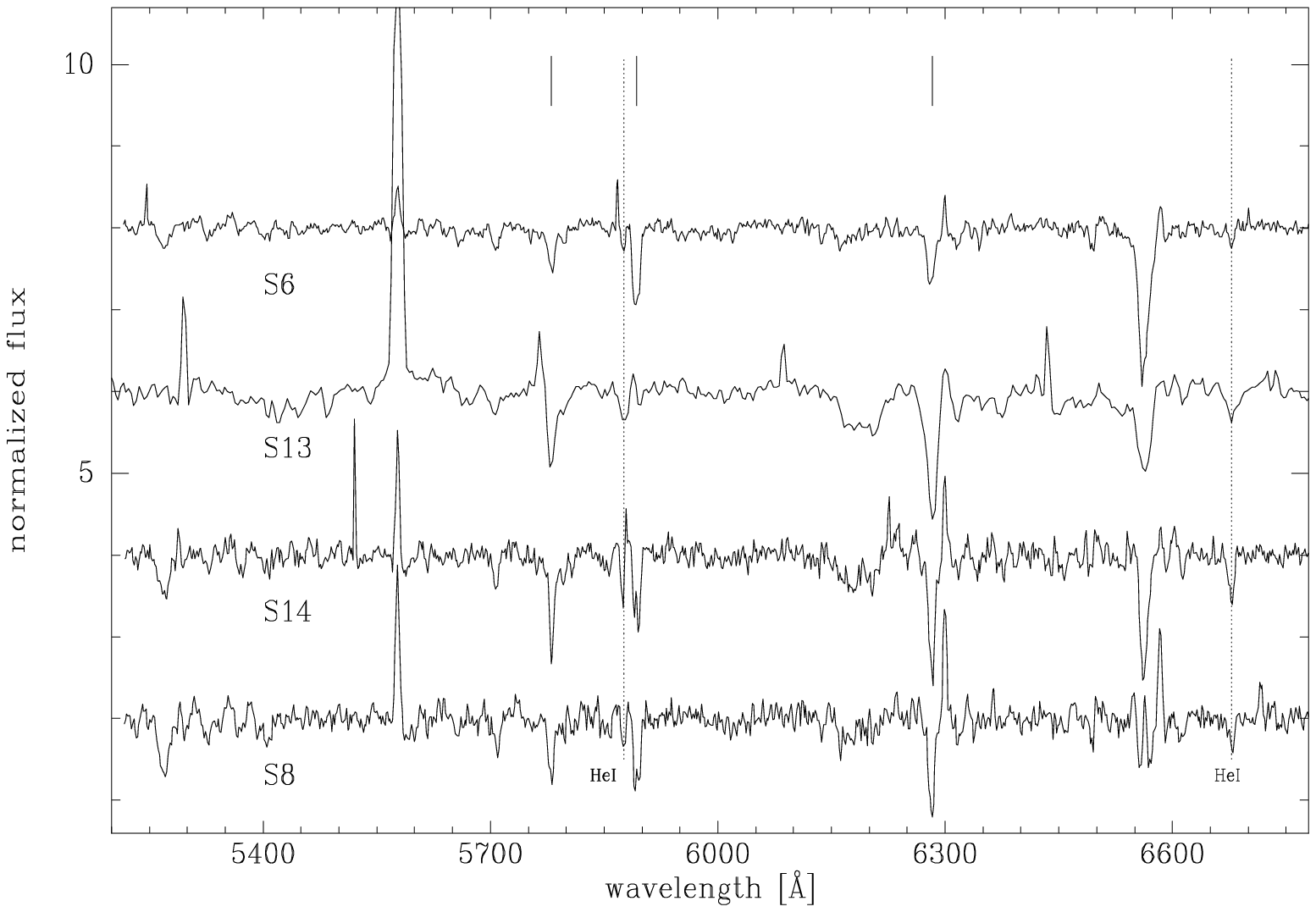}
\caption{Optical low resolution spectra of O and B stars,
 upper and lower panel, respectively.
All detected stellar lines are indicated (dotted line),
short solid lines indicate interstellar absorption.}
\label{esp-star}
\end{figure}

\begin{table*}
\caption{N-IR photometric and optical spectral information from the observed stars. } 
\label{starsTable}
\centering
\begin{tabular}{l c c c c c c c}
\hline\hline
 Star &  2MASS designation & J & H & Ks & Main optical {\bf $\dagger$} & EW$_{5893}$ &  S/N\\
      &                & (mag) & (mag) & (mag) & feature      & (\AA)    &            \\
\hline
S1   & 18252783-1317114  & 11.224 & 10.223   & 9.774     & --                &  -- & 15 \\
S2   & 18252859-1316565  & 11.672 & 11.224   & 11.028    & Balmer lines                &  -- & 30 \\
S3   & 18253221-1316438  &  9.005 & 8.395    & 8.093     & HeII absorption lines         &  2.2 & 30 \\
S4   & 18251888-1312139  & 10.739 & 9.592    & 9.088     & No Balmer lines  &  1.5 & 40 \\
S5   & 18251808-1309427  & 10.293 & 9.770    & 9.472     & HeII absorption lines         &  1.5 & 30 \\
S6   & 18251894-1308045  & 10.811 & 10.709   & 10.647    & HeI absorption lines         &  0.2 &  50\\
S7   & 18251943-1311340  & 10.191 & 9.580    & 9.389     & No Balmer lines      &  8.5 & 40 \\
S8   & 18245470-1317125  & 10.990 & 10.459   & 10.238    & HeI absorption lines          &  1.0 & 70 \\
S9   & 18245468-1314584  & 11.576 & 11.018   & 10.643    & HeI absorption lines(?) &  0.8 & 50 \\
S10{\bf*} & 18250115-1315490  & 9.460  & 8.757    & 8.552     & Balmer lines                &  2.7 & 30 \\
S11  & 18250674-1314526  & 10.847 & 10.613   & 10.597    & Strong Balmer lines               &  1.2 & 170  \\
S12  & 18250312-1314462  & 11.330 & 11.134   & 11.071    & Strong Balmer lines               &  1.2 & 90  \\
S13{\bf*} & 18250606-1316423  & 9.707  & 9.119    & 8.869     & HeI absorption lines          &  1.0 & 50 \\
S14{\bf*} & 18250579-1316349  & 10.750 & 10.162   & 9.906     & HeI absorption lines          &  0.8 & 110 \\
S15  & 18250290-1309385  & 8.046  & 7.509    & 7.158     & HeII absorption lines         &  2.3  & 30 \\
S16{\bf*} & 18244104-1309573  & 10.253 & 9.799    & 9.443     & Strong Balmer lines      &  --   & 70     \\
\hline
\multicolumn{7}{l}{Note: The {\bf*} in the sources means that no nebular emission at H$_\alpha$ is detected near
the star.}\\ 
\multicolumn{7}{l}{ {\bf $\dagger$} HeI absorption lines(?): the spectrum shows some hints of HeI absorption lines.} \\
\multicolumn{7}{l}{~ No Balmer lines: Balmer lines are not detected, suggesting to be a late spectral type.}
\end{tabular}
\end{table*}

\subsubsection{Stellar content}

Table \ref{starsTable} presents the observed stars with their 2MASS designation and JHKs photometry (Columns 1 to 5). 
In Columns 6 and 7 we include the main optical spectroscopic feature obtained from the spectral analysis and the Equivalent Width (EW) of 
the diffuse interstellar band (DIB) at 5893 \AA, respectively. Finally, the S/N of the spectra is presented in Column 8.
In Column 6, we remark if absorptions lines of HeI and HeII are observed in the stellar spectrum, which is the 
most important feature for our analysis. Following \citet{walborn90} we can infer that the stars with HeII absorption
lines are O-type stars (S3, S5, and S15), while those sources with only HeI absorption lines (sources S6, S8, S13 and S14) should be B0-5 stars. 
The spectra of these sources are shown in Figure \ref{esp-star}. Particularly interesting is the spectrum of source S15, which shows emission of NIII, 
indicating a star of population I or III \citep{walborn90}. From the I(4541)/I(4471) ratio we can infer that S15 may be an O8-9 star.
In the case of source S9, the spectrum (not presented in Figure \ref{esp-star}) shows some hints of HeI absorption lines, but it is 
not possible to confirm it due to the low S/N.

For sources S3, S5, and S15, the O-types stars, we estimated their radial velocities using the spectra obtained with the 600 grooves mm$^{-1}$ grating.
The radial velocities were estimated by averaging the values obtained from the HeII and HeI absorption lines (at wavelengths of 5411.52, 5875.62 and 6678.15 \AA).
The obtained velocities with respect to the local standard of rest, are: v$_{\rm LSR} \sim$ (53, 52, and 32) $\pm$ 23 \ks for S3, S5 and S15, respectively. 
These velocities, within the large error bar, are roughly in agreement with the molecular gas velocity range and with the velocities of the HII regions 
(see Sections \ref{present} and \ref{ism}). 

One important result is that we confirm the presence of a not catalogued HII region (the region indicated with an interrogation mark in Figure 
\ref{presentfig}), 
from which we very likely identified its exciting O-type star (source S15).

As mentioned above, \citet{watson08} suggested some stars to be the possible
exciting sources of N21. Two of those stars are included in our sample: S1 and S3 (IN21-2 and IN21-1, respectively in \citealt{watson08}).
According to these authors, our source S3 (IN21-1 in their work) is the best candidate star for producing N21. They point out that spectroscopic
observations suggest that it should be a late-O supergiant star, which is in agreement with our observations.
In the case of source S1, we could not observe any significant stellar feature due to the low S/N ratio, 
being impossible to conclude anything on this source.

\subsection{X-ray point sources}

\begin{figure}
\centering
\includegraphics[width=9cm]{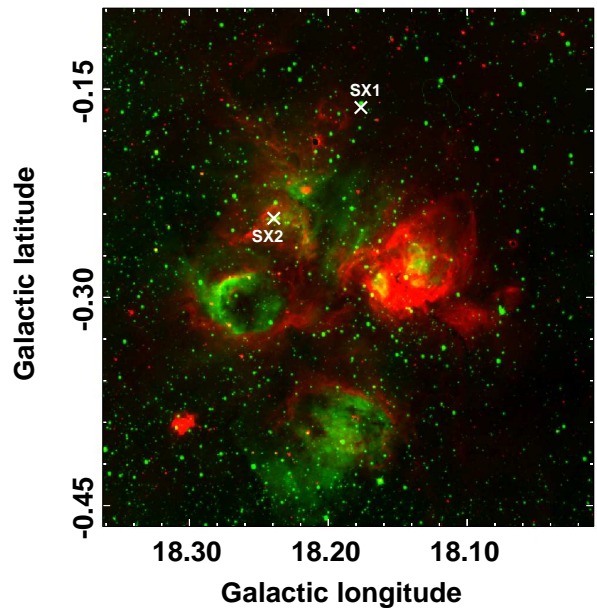}
\caption{Same as Figure \ref{exciting} with the two X-ray point sources found in the Chandra Source Catalog.}
\label{starsX}
\end{figure}

Since first X-ray observations with the Einstein, ROSAT and ASCA satellites, it is well
established that hot massive stars are intense X-ray emitters.
We look for X-ray point sources in the Chandra Source Catalog \citep{evans10} toward the studied region. 
We found two sources: CXO J182438.1-131039 and CXO J182502.8-130933 (hereafter SX1 and SX2, respectively), which are shown in Figure
 \ref{starsX}. It is important
to note that source SX2 coincides with the above analysed source S15, a very likely O-type star which is exciting
an uncatalogued HII region, while SX1 does not coincide with any source analysed at the optical wavelengths and 
lies in projection over the SNR shell close to the HII region G18.197-00.181.

These sources were observed 
in two observation runs. One was with the HRC on 2012 August 20 (Obs.Id. 9006). 
However, the total exposure of this observation was too short ($\sim$1.1 ksec) to perform a reliable analysis. 
The other observation run was obtained with the ACIS detector on 2012 June 19 (Obs.Id. 11883)
with a total exposure time of 10.0 ksec. The data was acquired in faint mode,
with six CCD turned on, two comprising the ACIS-I array [2,3], plus CCDs
[5,6,7,8] of ACIS-S. However, the data from two CCDs of ACIS-I are not used
in the followings because of the degraded point spread function and
reduced effective area.

\subsubsection{Analysis}
\label{secAnaly}

To analyse the data and produce a level 2 event file, we processed the
level 1 event list using the CIAO 4.4 and CALDB 4.5.3 set of calibration
file.
We retain events with grades = 0,2,3,4,6 and status = 0, and assume a
constant background flux due to the lack of flaring background events. To
improve the sensitivity to faint sources, we filtered out events outside
[500:8000] eV band. We generated a source image and a congruent exposure
map (assuming a monochromatic spectrum, kT=2.0 keV) following standard CIAO
threads\footnote{http://asc.harvard.edu/ciao/threads/expmap\_acis\_single/}.

 Unfortunately, in the observation Id. 11883 the source SX2 lies at the gap between ACIS-I and ACIS-S chips. However,
we used the short observation Id. 9006 in order to estimate an upper limit X-ray emission level 
for this source. We were able to detect 
SX1 as an X-ray source with a total of $\sim$ 80 photons, and for SX2, we
obtained a very marginal detection with a total of 5 background corrected
photons.

It can be noticed that we do not observe diffuse emission from the SNR G18.1-0.1. As mention above, 
this SNR  was previously detected in X-rays with the ASCA satellite in the 0.7-10 keV energy range \citep{sugizaki}. 
These authors converted X-ray photons to energies using a single absorbed 
power-law model to obtain a flux of $5.4 \times 10^{-13}$ erg cm$^{-2}$  s$^{-1}$. However, taking into account the ASCA low spatial 
resolution (40 arcmin) and 
the SNR size (8 arcmin), this flux must be spanned over this area. By rescaling this flux at the Chandra spatial resolution
we computed the diffuse X-ray emission level of the SNR. We used  the Pimms\footnote{http://heasarc.gsfc.nasa.gov/Tools/w3pimms.html} 
software to obtain an upper limit flux density of about 
$1.9 \times 10^{-14}$ erg cm$^{-2}$ s$^{-1}$ arcmin$^{-2}$, which is below the detection threshold of this ACIS-S observation.

\begin{figure}
\centering
\includegraphics[width=6cm,angle=-90]{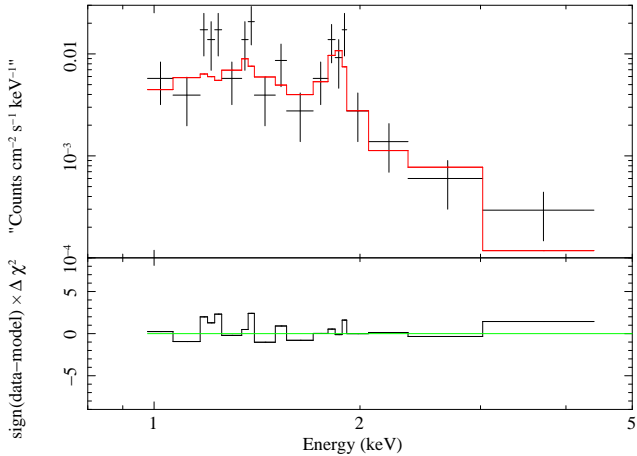}
\caption{X-ray spectrum obtained toward source SX1. The spectrum was binned to reach a minimum of four and nine counts per
channel. The red line shows the best fit model. The broad observed emission lines (line blending effect) are produced due to the low resolution of the spectra.
The reduced chi-square is 1.05 for 16 degree of freedom.}
\label{spectra}
\end{figure}

To characterize the hot plasma responsible of the X-ray
emission of  source SX1, and to estimate the intrinsic X-ray luminosity, we
analysed the ACIS-S spectra. We extracted source photons by using a circular region
of radii 2.9 arcmin.  A total of 87 X-ray photons were acquired from source SX1 for a total of 10 ksec with a median energy of
1.53 keV.
The spectral fitting model consists of a combination of absorption and emission models.
Absorption {\sc wabs} model was used to account for the action of the circumstellar
medium, while we assumed emission by a thermal plasma, in collisional ionization
equilibrium, as modeled by the APEC code \citep{smith01}.
Elemental abundances are not easily constrained with low-statistic spectra, but it
was left free during the fitting procedure.
Spectral parameter are shown in Table \ref{tableX}, while Figure \ref{spectra} shows
the observed SX1 spectrum and the fitted solution.
It is clear that some spectral emission lines appear blended, being the most intense
from SiXIII (1.864 keV).

\begin{table}
\caption{X-ray parameters of source SX1.}
\label{tableX}
\begin{tabular}{ll}
\hline
Parameter & SX1 \\
\hline
N$_{\rm H}$ [cm$^{-2}$]& 1.82($\pm$0.32)$\times$10$^{22}$ \\      
kT [keV]        & 0.59 $\pm$ 0.12  \\      
Abundance       & 1.62 $\pm$ 0.93   \\      
Norm.           & 2.5($\pm$ 1.48)$\times$10$^{-4}$  \\     
Flux [cgs]      & 3.94$\times$10$^{-14}$   \\
Unabsorbed Flux [cgs] &1.18$\times$10$^{-12}$      \\
L$_{\rm x}$ [erg/s]   & 2.25$\times$10$^{33}$     \\      
\hline
\end{tabular}
\end{table}

In the 0.5-8.0 keV range, we calculated an X-ray flux of
3.94$\times$10$^{-14}$  erg\,cm$^{-2}$\,s$^{-1}$~for SX1,
while the corrected absorbed X-ray flux is 1.18$\times$10$^{-12}$
erg\,cm$^{-2}$\,s$^{-1}$.
Assuming a distance of 4 kpc, the unabsorbed X-ray luminosity is L$_{\rm
x}$$\sim  2.25 \times 10^{33}$ erg s$^{-1}$, which is a typical emission level of an early O-type star (O3-O5).
The estimated upper limit X-ray emission level for source SX2 suggests that it also could be an O-type star, however longer 
exposure observations are needed in order to confirm it.
In conclusion, the X-ray analysis strongly suggests the presence of other early O-type star in the region (SX1) and encourage to obtain longer X-ray observations
toward the region in order to study, in the high energy regime, the OB-type stars content.

\section{The HII region/SNR complex surroundings: the big picture}
\label{larger}

\begin{table*}
\caption{Main physical parameters of the big molecular shell.}
\centering
\begin{tabular}{cccccc}
\hline
R$_{eff}$ & V$_{sys}$ & V$_{exp}$ = $\Delta$v/2 & Mass & E$_{kin}$ & t$_{dyn}$ \\
pc & \k & \k &  $\times 10^5$ \msol & $\times 10^{48}$[erg] & Myr \\
\hline
24.5$\pm$1.3 & 53.1$\pm$0.2 & 2.0$\pm$0.2& 2.0$\pm$1.1 & 9.0$\pm$0.6 & 7.0$\pm$2.7\\
\hline
\label{para}
\end{tabular}
\end{table*}

\begin{figure}
\centering
\includegraphics[width=9.5cm]{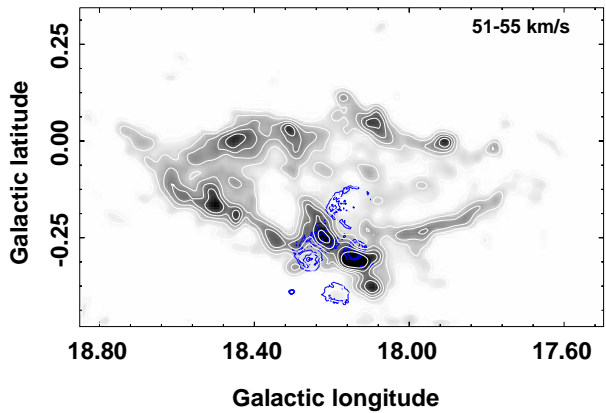}
\caption{Large scale ISM: \3 J=1--0 emission integrated between 51 and 55 \ks with white contours
of levels of 1.2, 1.8, 3.0, 4.0 , 5.8, and 7.5 K \k. The blue contours correspond to the radio continuum emission at 20 cm,
showing the SNR and the HII regions positions.}
\label{large}
\end{figure}

In this section we analyse the HII region/SNR complex surroudings in a larger spatial scale.
A detailed inspection of the whole \3 data cube at larger scale reveals
that the HII region/SNR complex lies in a border of a big molecular
shell that peaks in the velocity interval of 51--55 \k. By inspecting the mid-IR emission in this large scale
area we did not find any possible counterpart to this molecular shell. 
Figure \ref{large} presents the \3 J=1--0 emission of a large region of about 1\fdg3 $\times$ 0\fdg8 integrated between 51 and 55 \k.
It is very interesting to note that the HII region/SNR complex is located southward a very likely molecular shell
with an elliptical shape with semi-axes of 30\m$\times$12\m, centered at $l =$ 18\fdg185, and $b = -$0\fdg104.
Assuming a distance of 4 kpc, the molecular shell has an extension of about 70 pc $\times$ 28 pc.
Following the standard nomenclature for Galactic
shells, we designate this new structure as GSH 18.2-0.1+53. Along the
mentioned velocity range, the CO shell is well defined as a large
structure, although, as in many other shells, the typical behaviour of
an expanding structure (i.e. a ring that starts as a cap, reaches its
greatest angular dimension at the systemic velocity and then shrinks
back to a cap) is not completely observed. 
Figure \ref{large} also shows that the shell is composed of several
molecular condensations among which are those associated with the HII
region/SNR complex. 

Following the same assumptions and equations as Section \ref{ism}
we derive an average H$_ {2}$ column density for the whole molecular shell.
Assuming a constant T$_{\rm ex}$ = 20 K over the shell and a
$\Delta v =  4$~\k, we obtain N(H$_{2}$) $\sim$ 5 
$\times 10^{21}$ cm$^{-2}$. Using Equation (\ref{eq4}) and performing the summation over all the 
observed positions within the 1.2 K \ks contour level (which define the shell boundary, see Figure \ref{large}) we obtain 
a total molecular mass of about 2 $\times 10^5$ \msol. 
Following \citet{wea77}, we estimate the kinetic energy stored in the shell and its age using 
$E_{kin} = 0.55 \times M_{shell} \times V_{exp}^2$ and $t_{dyn} = 0.6
R_{eff} / V_{exp}$, respetively, in which is considered a simple model that describes the expansion 
of a shell created by a continuous injection of mechanical energy, where $R_{eff}$ is the effective radius.
Table \ref{para} summarizes the main physical parameters of GSH 18.2-0.1+53.

The origin of most of these expanding structures would be related to
the action of massive stars (OB-type stars) and their descendants (WR
stars), which strongly disturb the environment that surrounds them,
first by the high rate of ionizing photons and strong winds, and
subsequently at the end of their lives, when they explode as
supernovae. In this context, we wonder for the origin of GSH
18.2-0.1+53. 
We have not found any OB star catalogued toward the interior of the shell. 
But of course we cannot discard that this shell has been originated by the action of massive stars. Our X-ray analysis 
suggests that SX1, which lies in the interior of the molecular shell, can be an O-type star, probably one of the 
responsibles, together with other undetected OB-type stars and even the SNR progenitor, in generate the shell.
Regarding to a SNR origin we consider two possible scenarios: i) the contribution of
the SNR G18.1-0.1, and ii) the action of an ancient SNR not observed at
the present day. Given that the effective action of SNR G18.1-0.1
is limited to the borders defined by the radio continuum shell, and considering that typical ages for SNRs are at most 10$^5$
yrs, we discard that G18.1-0.1 has originated the
big shell by itself. However, as mentioned above, its massive progenitor could have contributed to generate it. 
By the other hand, considering the scenario where
the origin of GSH 18.2-0.1+53 is attributed to an ancient SNR that
would have injected to the ISM a mechanical energy of about $10^{51}$
erg, from which the shell still preserves about 1\% of that energy, it could be a similar 
case as was proposed for the surroundings of the SNR G349.7+0.2 \citep{reynoso01}. 
In that case it was found that the SNR, some IRAS sources and an UCHII region lie in a border of an extended molecular shell, 
suggested to be the fossil remains of an ancient SN explosion.
Summarizing, the origin of GSH 18.2-0.1+53 still remains uncertain and it deserves a deeper study, e.g. new observations toward
its interior in order to look for OB-type stars.

\section{Summary}
\label{summ}

The SNR G18.1-0.1 is located, along the plane of the sky, close to several HII regions (infrared dust bubbles N21 and N22, and the 
HII regions G018.149-00.283 and G18.197-00.181). Taking into account
the catalogued distances and/or systemic velocities of these estructures, and our analysis of the molecular gas related
to them, we suggest that all of these objects belong to a same complex at a distance of about 4 kpc. This could be a clear observational evidence 
supporting that massive stars born predominantly within dense cores of giant molecular clouds and evolve in clusters, which can generates
several neighboring HII regions and SNRs.
The main results of this work can be summarised as follows: \\

(1) Analysing the $^{13}$CO J=1--0 emission toward the HII region/SNR complex we found several molecular structures very likely related to
the sources within the velocity range between 39 and 59 \k. The molecular gas morphology suggests 
that it is very likely being swept and shaped by the expansion of the HII regions, mainly by the infrared dust bubbles 
N21 and N22, and the HII region G018.149-00.283. We obtained a molecular mass for the whole structure related to the complex 
of about  $10^{5}$ \msol. We estimated a density of about $9 \times 10^{3}$ cm$^{-3}$ for the densest portion of this molecular structre.

(2) Using spectroscopic optical observations obtained with the 2.15 m telescope at CASLEO, Argentina, we searched for OB-type stars in order to 
look for sources responsible of ionizing the gas in the HII regions.
We discovered three O-type stars verly likely exciting the bubbles N21 and N22, and an uncatalogued HII region northward bubble N22, respectively. Thus, 
we confirm the presence of an unknown HII region. Additionally, we discovered four B0-5 stars, one toward the bubble N22 and the others within
the HII region G018.149-00.283.

(3) Taking into account that hot massive stars usually are X-ray emmiters, we looked for X-ray point sources in the Chandra Source Catalog. 
We found two sources, one of them coincides with the above mentioned star
that is exciting the uncatalogued HII region. Estimating an upper limit for its X-ray emission level we can only suggest that it may be an O-type star. 
On the other side, the X-ray analysis of the other source, which lies over the SNR shell, strongly suggests that it is an early O-type star (O3-O5).

(4) By inspecting the ISM surrounding the HII region/SNR complex in a large scale, we discovered a big molecular
shell peaking between 51 and 55 \k. Assuming a distance of 4
kpc, this shell has an extension of about 70 pc $\times$ 28 pc, and the analysed HII region/SNR complex appears to be 
located in its southern border.

\section*{Acknowledgments}

We wish to thank the anonymous referee whose comments and suggestions have helped to considerably improve this paper.
S.P., W.W., M.O. and J.F.A.C. are members of the {\sl Carrera del 
investigador cient\'\i fico} of CONICET, Argentina. A.P. is a postdoctoral fellow of CONICET, Argentina.
This work was partially supported by Argentina grants awarded by UBA (UBACyT), CONICET and ANPCYT. The CCD and data acquisition 
system at CASLEO has been financed by R. M. Rich trough U.S. NSF grant AST-90-15827.

\label{lastpage}
\end{document}